\documentclass[preprint,12pt]{elsarticle}
\journal{Few Body Systems}
\begin{document}

\begin{frontmatter}
\title{Improvement of multichannel amplitudes for the pion-pion scattering 
using the dispersion relations \footnote{Presented at the 20th International 
IUPAP Conference on Few-Body Problems in Physics, 20 - 25 August, 
2012, Fukuoka, Japan}}

\author{Petr Byd\v{z}ovsk\'y}
\address{Nuclear Physics Institute, ASCR, \v{R}e\v{z}/Prague, Czech Republic}
\author{Robert Kami\'nski}
\address{Institute of Nuclear Physics, PAN, Krakow, Poland}

\begin{abstract}
The multichannel S- and P-wave amplitudes for the $\pi\pi$ 
scattering, constructed requiring analyticity and unitarity of the 
S-matrix and using the uniformization procedure, are elaborated using  
the dispersion relations with imposed crossing symmetry condition. 
The amplitudes are modified in the low-energy 
region to improve their consistency with experimental data and the 
dispersion relations. Agreement with data is achieved for both 
amplitudes from the threshold up to 1.8 GeV and with dispersion relations 
up to 1.1 GeV. Consequences of the applied modifications, e.g. changes 
of the S-wave lowest-pole positions, are presented.
\end{abstract}

\end{frontmatter}

\section{Introduction}
\label{intro}
A model independent analysis of the $\pi\pi$ 
scattering is important 
tool in getting information about the spectrum of light mesons. 
A reliable description of the process is therefore desirable to 
allow us to learn more on parameters of the mesons. 

The phenomenological multichannel amplitudes for the S and P 
waves in the $\pi\pi$ scattering were constructed without any 
specific assumptions about dynamics of the process, only requiring 
analyticity and unitarity of the S-matrix and applying the 
uniformization procedure~\cite{YuSu2010}. 
This procedure can be applied exactly in the two-channel case. 
However, in the three-channel case, simplifying approximations 
have to be done resulting in a very poor description of experimental 
data in the threshold region~\cite{YuSu2010}. 

The crossing symmetry condition, which relates the S and P waves 
and which is important below the inelastic threshold, was not taken 
into account in the construction of these amplitudes~\cite{YuSu2010}. 
Since the crossing symmetry is properly included in the Roy-like 
dispersion relations~\cite{DR1}, it is possible and desired 
to improve the low-energy behavior of the three-channel 
amplitudes~\cite{YuSu2010} and to check their consistency with 
the dispersion relations (DR). 

In the multichannel uniformizing (MI) approach a heavy and broad 
$\sigma$ meson is predicted, $m = 829\pm10$ MeV and 
$\Gamma = 1108\pm22$ MeV~\cite{YuSu2010}, in disagreement 
(by many standard deviations) with results from DR~\cite{DR2} 
and values recomended by Particle Data Group~\cite{PDG}. 
It is therefore interesting to 
show how much the modifications of the three-channel amplitudes 
affect positions of poles connected with the $\sigma$ meson.  
The analysis can also contribute to disclosing reasons of differences 
between some results from the MI and DR approaches. 

In this note we present an example of the use of dispersion relations 
for improving the low-energy behavior of a phenomenological 
three-channel S- and P-wave  $\pi\pi$ amplitudes and imposing  
the crossing symmetry condition on the amplitudes below 1.1~GeV.

\section{Multichannel amplitudes and dispersion relations}
\label{sec:1}
Two channels coupled to the $\pi\pi$ channel, $K\bar{K}$ 
and $\eta\eta'$ for S wave and $\rho 2\pi$ and $\rho\sigma$ 
for P wave, were explicitly considered in construction of 
the three-channel amplitudes~\cite{YuSu2010}. 
The eight-sheeted Riemann surface was transformed into 
a uniformization plane using a uniformizing variable $w$ in which  
the left-hand branch point connected with the crossed channels 
was not taken into account. Therefore the crossing symmetry condition 
was not considered in the construction. 
A contribution of the left-hand cut was included in the background 
part of the amplitude. 
Note, that in Ref.~\cite{YuSu2012} 
the left-hand branch point in $w$ was already included in the S-wave 
analysis. 

In the uniformization plane an influence of the $\pi\pi$-threshold 
branching point was neglected, keeping however the unitarity on the 
$\pi\pi$ cut, which was a necessary approximation in the three-channel 
case~\cite{YuSu2010}.
We got, therefore, a four-sheeted model of the initial Riemann surface 
in which the near-threshold data are not described properly. 
Note that in the two-channel case this approximation is not needed and 
the threshold data are described correctly~\cite{arxiv2012}. 

The resonance part of the matrix element $S_{11}$ of the S-matrix 
is generated by clusters of complex-conjugate poles and zeros 
on the Riemann surface, which represent resonances~\cite{YuSu2010}.
For example, the $f_0(500)$ (formerly was $f_0(600)$) resonance 
is represented by a cluster 
which possesses zero only in the $S_{11}$ matrix element on the 
physical sheet. Location of poles on the unphysical sheets is given 
by the analytic continuation of the matrix elements~\cite{YuSu2012}. 
The background and resonant parts of the S-matrix are separated 
and expressed via the Le Couteur--Newton relations with the Jost matrix 
determinant $d(w)$~\cite{YuSu2010}  
\begin{equation} 
S_{11} = S^{bgr}_{11} S^{res}_{11} = 
\frac{d_{bgr}(-k_1,k_2,k_3)}{d_{bgr}(k_1,k_2,k_3)} 
\frac{d_{res}^* (-w^*)}{d_{res}(w)}\,, 
\end{equation}
where $k_j$ are the channel momenta. The resonant part is    
$d_{res}(w)=w^{-\frac{M}{2}}\prod (w+w_{r}^*)\,,$   
where the product includes all zeros $w_r$ of the chosen resonances 
and $M$ is a number of resonance zeros. 
The background part is modeled via complex energy-dependent phases 
$\alpha_j(s)$, $j$=1,2,3  representing mainly an influence of other 
channels and the neglected left-hand cut: 
$d_{bgr}(k_j)=\exp\left[{\rm -i}\sum \alpha_j(s)\right]$. 
The resonance zeros $w_r$ and background parameters 
were obtained from fitting the phase shifts and inelasticity parameters 
in the assumed channels to experimental data~\cite{YuSu2010}.

The once-subtracted dispersion relations with imposed crossing symmetry 
condition for S ($l$=0, $I$=0) and P ($l$=1, $I$=1) waves read as
\begin{equation}
\ \ \ {\rm Re}f_l^I(s)^{out} = ST_l^I +  
\sum_{I'=0}^{2}\sum_{l'=0}^{3}{\rm vp}\int_{4m_\pi^2}^{s'_{max}} 
ds'\,K_{ll'}^{II'}(s,s')\, {\rm Im}f_{l'}^{I'}(s')^{in}\ 
+\ d_l^I(s)\,,
\label{DR}
\end{equation}
where $ST_l^I$, $K_{ll'}^{II'}(s,s')$ and $d_l^I(s)$ are the subtracting, 
kernel and driving terms, respectively~\cite{DR1}. $f_l^I(s)^{out}$ 
and $f_{l'}^{I'}(s')^{in}$ are the output and input amplitudes. 
The difference between Re$f_l^I(s)^{out}$ and Re$f_l^I(s)^{in}$ 
demonstrates a consistency of the amplitudes with the dispersion relations 
(i.e. with crossing symmetry). 
The smaller the difference the better consistency, see the last term in 
eq.~(\ref{chi2}) below. The summation includes also D and F waves 
described by phenomenological expressions \cite{DR1}.

\section{Method of improvement of the amplitudes at low energies}
\label{sec:2}
The near threshold behavior of the S- and P-wave amplitudes   
is determined by a generalized expansion in powers of the pion momentum 
$k$ = $\sqrt{s/4-m_\pi^2}$~\cite{DR1}
\begin{equation}
{\rm Re}f_l^I(s)=\frac{\sqrt{s}}{4k}\,\sin 2\delta_l^{(I)}=
      m_\pi k^{2l}[a_l^I + b_l^I\,k^2 + c_l^I\,k^4 + 
      d_l^I\,k^6 + {\cal O}(k^8)]\,.
\label{expansion}
\end{equation}
The amplitudes given by this expansion are matched with those for 
higher energies from~\cite{YuSu2010} (the {\it original} amplitudes) 
at energies $s_{0l}$ fitted to data. In eq. (\ref{expansion}), $a_l^I$ 
is the scattering length and $b_l^I$ is the slope parameter fixed at values:  
$a_0^0=0.22\,m_\pi^{-1}$, $b_0^0=0.278\,m_\pi^{-3}$, $a_1^1=0.0381\,m_\pi^{-3}$, 
and $b_1^1=0.00523\,m_\pi^{-5}$~\cite{DR1}.    
$c_l^I$ and $d_l^I$ are calculated from the continuity conditions 
for the phase shift and its first derivative at the matching energies $s_{0l}$. 
The low-energy corrected original amplitudes are denoted by {\it extended} amplitudes. 
Above $s_{0l}$ the {\it original} and {\it extended} amplitudes 
are equivalent.

Parameters of the extended amplitudes, which strongly influence the low-energy 
behavior of the amplitudes, were optimized (re-fitted) to fit the experimental 
data and to achieve a better consistency with the dispersion relations, minimizing
\begin{equation}
\chi^2 = \sum_{i}\left(\frac{\delta_i^{exp}-\delta_i^{th}}{\Delta\delta_i^{exp}}\right)^2
+\sum_{i}\left(\frac{\eta_i^{exp}-\eta_i^{th}}{\Delta\eta_i^{exp}}\right)^2
+\sum_{i}\left(\frac{{\rm Re}f_i^{out}-{\rm Re}f_i^{in}}{\Delta_{DR}}\right)^2\,,
\label{chi2}
\end{equation}
where $\delta_i$ and $\eta_i$ are experimental and calculated values of 
the phase-shift and inelasticity parameter in the assumed channels of 
the S and P waves. The summation therefore runs also over the channels and 
partial waves. The scale parameter $\Delta_{DR}$ 
makes a reasonable weight of the DR contribution to $\chi^2$. 
Note, that the last term in (\ref{chi2}) provides a coupling between 
the S and P waves which would be otherwise independent in the analysis. 

The re-fitted parameters are zeros of the lowest poles, 
$f_0(500)$, $f_0(980)$, and $\rho(770)$, the matching energies $s_{00}$ and 
$s_{01}$, and the background parameters in the $\pi\pi$ channel. 
Experimental data used in this analysis are from Ref.~\cite{YuSu2010} 
supplemented near the threshold with phases from the dispersive analysis~\cite{DR1} 
and data from the NA48 Collaboration.

\section{Results}
\label{sec:3}
Applying the modifications we got new S- and P- wave amplitudes 
which describe very well the experimental data on the $\pi\pi$ 
scattering from the threshold up to 1.8 GeV.   
The threshold expansion (\ref{expansion}) provided a reasonable agreement 
with data, $\chi^2/{n.d.f.}$ = 2.36 for the extended amplitudes, but the 
re-fitting of parameters still significantly improved the result, 
$\chi^2/{n.d.f.}$ = 1.29 for the re-fitted amplitudes. The biggest 
improvement was for the DR contribution, the last term in eq. (\ref{chi2})   
changed from 571 to 66, which suggests a significant improvement of consistency 
of the amplitudes with the Roy-like dispersion relations. The re-fitted amplitudes 
provide also proper values of the phase shifts and inelasticity parameters 
in the assumed coupled channels as the original amplitudes. 

Positions of poles changed strongly for the $f_0$(500) resonance, e.g. 
on sheet II the pole shifted from $617-i\,554$ MeV for the original 
amplitude to $474-i\,298$ MeV for the re-fitted one which results in 
a reduction of the $\sigma$ meson mass, 829 MeV $\rightarrow$ 560 MeV 
and width, 1108 MeV $\rightarrow$ 596 MeV. 
Note that the new pole position accords well with the result 
from the analysis based on ChPT and Roy-like equations, 
$(441^{+16}_{-8}-i(272^{+9}_{-12.5})$~\cite{Gilberto}.   
The poles of $f_0(980)$ shifted slightly, e.g. on sheet II  
from $1013-i\,31$ MeV to $1000-i\,22$ MeV which makes the new mass, 1000 MeV, 
and width, 45 MeV, more consistent with the values suggested by Particle Data 
Group, $990\pm20$ MeV and $50^{+20}_{-12}$ MeV, respectively~\cite{PDG}.   
The poles of $\rho(770)$ moved up  by less than 1\%. 

Re-fitted values of the background parameters are small suggesting that 
important part of dynamics is included in the resonant part of the S-matrix. 
However, in the S-wave the background phase shift became negative starting at 
the $\pi\pi$ threshold ($a_{11}= -0.091$) which seems to be necessary for 
a good description of data. 
  
To summarize, agreement of the phase-shifts with low-energy data was 
improved  for the new re-fitted S- and P-wave $\pi\pi-$scattering 
amplitudes. 
The amplitudes are calculated with the scattering lengths and 
slope (effective-range) parameters consistent with results of calculations 
based on ChPT and DR.   
Consistency of the three-channel amplitudes with the dispersion 
relations was improved significantly from the threshold up to 1.1 GeV 
which means that the amplitudes better fulfill the crossing symmetry 
condition. 
The lowest pole in S wave is shifted to lower energy and nearer to the 
real axis which results in smaller values of the mass and width for 
the $\sigma$ meson. 
However, the S-wave background phase-shift is negative beginning from the $\pi\pi$ 
threshold.

\noindent ACKNOWLEDGMENT:
We are grateful to Gilberto Colangelo for an inspiring discussion 
and interest in our work. 
We thanks the Grant Agency of the Czech Republic, grant P203/12/2116.
This work has been partly supported by the Polish Ministry of Science and
Higher Education (grant No N N202 101 368).

\end{document}